\documentclass[twocolumn,aps,prl,superscriptaddress]{revtex4-2}

\usepackage[colorlinks=true,allcolors=blue]{hyperref}
\usepackage{array}
\usepackage{amsmath}
\usepackage{amssymb}
\usepackage{graphicx}
\usepackage{babel}
\usepackage{braket}
\usepackage{mathrsfs}
\usepackage{amsfonts}
\usepackage{epstopdf}
\usepackage{multirow}
\usepackage{color}
\usepackage{bm}
\usepackage{amsthm}

\usepackage{xcolor}
\usepackage{tikz,fp}
\usepackage{tikz-cd}
\usepackage{enumitem}

\usetikzlibrary{arrows}
\usetikzlibrary{intersections}
\usetikzlibrary{shapes.geometric}
\usetikzlibrary{decorations.pathmorphing, patterns,shapes,fixedpointarithmetic}
\usetikzlibrary{decorations.markings}

\newcommand{\tr}{\mathrm{tr}}

\renewcommand{\leq}{\leqslant}

\begin{document}

\title{Reviving the Lieb--Schultz--Mattis Theorem in Open Quantum Systems}
\author{Yi-Neng Zhou}
\thanks{These two authors contributed equally.}
\affiliation{Institute for Advanced Study, Tsinghua University, Beijing 100084, China}
\author{Xingyu Li}
\thanks{These two authors contributed equally.}
\affiliation{Institute for Advanced Study, Tsinghua University, Beijing 100084, China}
\author{Hui Zhai}
\affiliation{Institute for Advanced Study, Tsinghua University, Beijing 100084, China}
\affiliation{Hefei National Laboratory, Hefei 230088, China}
\author{Chengshu Li}
\email{lichengshu272@gmail.com}
\affiliation{Institute for Advanced Study, Tsinghua University, Beijing 100084, China}
\author{Yingfei Gu}
\email{guyingfei@tsinghua.edu.cn}
\affiliation{Institute for Advanced Study, Tsinghua University, Beijing 100084, China}
\date{October 2, 2023}

\begin{abstract}

In closed systems, the celebrated Lieb--Schultz--Mattis (LSM) theorem states that a one-dimensional locally interacting half-integer spin chain with translation and spin rotation symmetry cannot have a 
non-degenerate gapped ground state. 
However, the applicability of this theorem is diminished when the system interacts with a bath and loses its energy conservation. 
In this letter, we propose that the LSM theorem can be revived in the entanglement Hamiltonian when the coupling to bath renders the system short-range correlated. 
Specifically, we argue that the entanglement spectrum cannot have a non-degenerate minimum, isolated by a gap from other states. 
We further support the results with numerical examples where a spin-$1/2$ system is coupled to another spin-$3/2$ chain serving as the bath. 
Compared with the original LSM theorem which primarily addresses UV--IR correspondence, our findings unveil that
the UV data and topological constraints also have a pivotal role in shaping the entanglement in open quantum many-body systems. 

\end{abstract}

\maketitle

\textit{Introduction.} 
Over sixty years ago, Lieb, Schultz, and Mattis proved their celebrated result, now known as the Lieb--Schultz--Mattis (LSM) theorem~\cite{Lieb1961,Affleck1986}. 
They considered a one-dimensional locally interacting half-integer spin chain with both translation and spin rotation symmetries and showed that this system cannot have a unique ground state with a finite excitation gap. In other words, either the ground states are degenerate, or the low energy spectrum is gapless. This theorem was later generalized to higher dimensions~\cite{Oshikawa2000, Hastings2004} where the ground state degeneracy can also originate from topological order, bringing out the connection between the LSM theorem and the search for the topological spin liquids~\cite{lee2008end,balents2010spin}. 
In the past decades, extensive studies on the LSM theorem have revealed its important role in understanding quantum magnetism and strongly correlated physics~\cite{affleck1989quantum,
chen2011classification,
Parameswaran13,
Zaletel_2015,
Watanabe_2015,
Cheng2016,
Cho2017,
Po_2017,
huang2017buildinga,
lu2017liebschultzmattis,
Yang_2018,
jian2018liebschultzmattis,
Metlitski2018,
fidkowski2018surface,
Cheng_2019,
Ogata2019,
Else2020,
Lu_2020,
jiang2021generalizeda,
ye2022topological,
gioiaNonzeroMomentumRequires2022,
zheng2022unconventional,
Tasaki2022,
Cheng2023}. Nevertheless, when a quantum system inevitably interacts with the environment, the concept of the energy spectrum is no longer well-defined. 
This raises an interesting issue of whether an LSM-type theorem can still be established for an open quantum system~\footnote{Recently an LSM-type theorem for open systems was proposed in arXiv 2305.16496 by Kawabata, Sohal, and Ryu. The authors considered the Lindbladian spectrum instead of the energy spectrum. We will provide an alternative proposal utilizing the entanglement spectrum.}.

In closed systems, the LSM theorem essentially states a deep ultraviolet--infrared (UV--IR) connection. 
The microscopic details (UV part), such as the spin per unit cell, non-trivially constrain the long wavelength (IR part) property of a many-body spin system. To be concrete, we consider the following spin twist operator
\begin{equation}
\hat{U}^\text{twist}=e^{i\frac{2\pi}{L}\sum_{j}j\hat{S}^z_j}, \label{twist}
\end{equation}
where $j$ labels site index and $L$ is the system size. The fact that $\hat{U}^\text{twist}$ anti-commutes with the translation operator $\hat{T}$ when acting on a spin-rotationally invariant state with half-integer spin per unit cell plays a central role in proving the LSM theorem. 
Note that this anti-commutating relation only relies on the intrinsic feature of the Hilbert space, independent of detailed Hamiltonian forms. 
Therefore, we anticipate a certain version of the LSM theorem can also be formulated in open systems.  

\begin{figure}[t]
    \centering
    \includegraphics[width=\columnwidth]{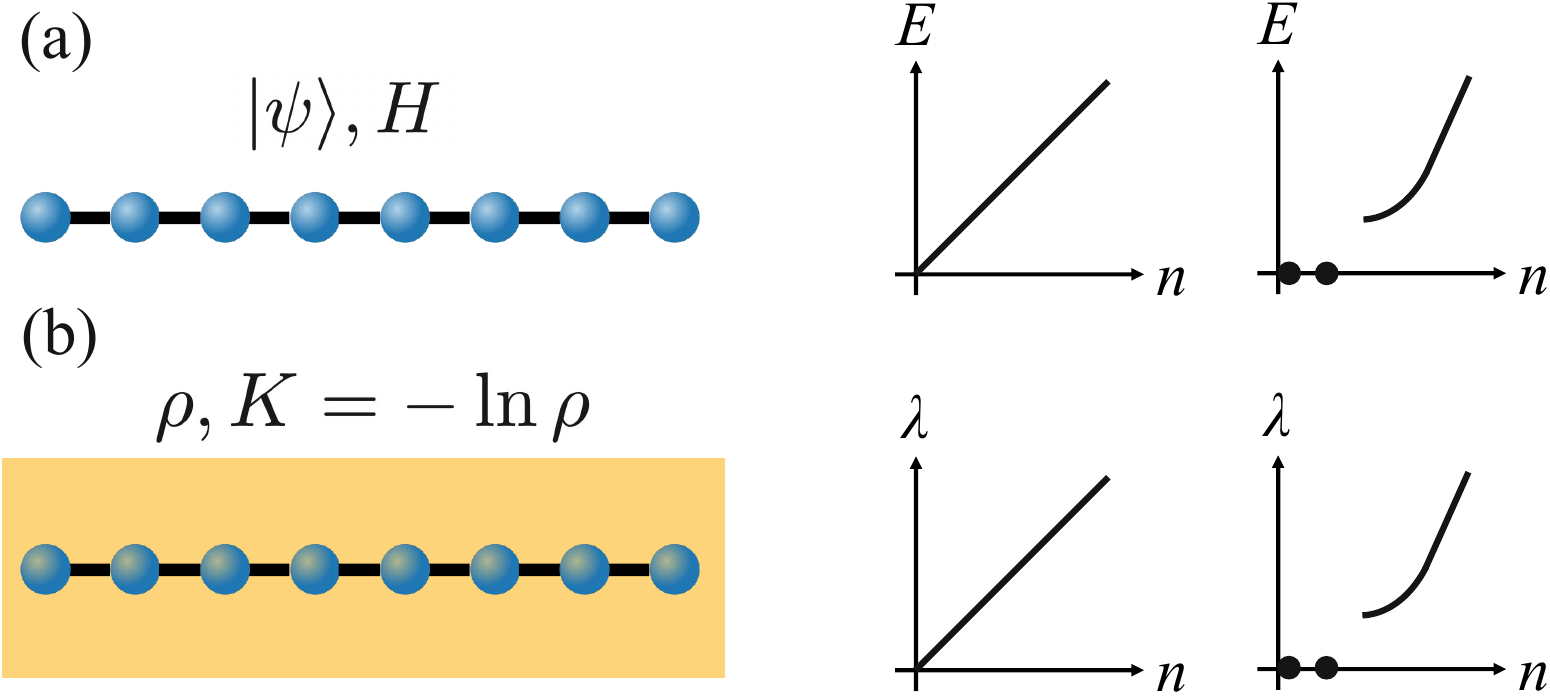}
    \caption{Comparing the LSM theorem in closed and open systems. (a) In closed systems, we consider the physical Hamiltonian $\hat{H}$ and the energy spectrum $E$. (b) In open systems, we consider the entanglement Hamiltonian $\hat{K}$ and the entanglement spectrum $\lambda$. These two spectra exhibit the same features, i.e. either gapless or degenerate ground states, when the LSM theorem holds.}
    \label{fig:ent_lsm}
\end{figure} 

When coupled to a bath, the system generically becomes a mixed state, described by a density matrix $\rho$. We write $\rho$ as $\rho=e^{-\hat{K}}$ where $\hat{K}$ is called the entanglement Hamiltonian whose spectrum is called the entanglement spectrum~\cite{Li2008}.
In this work, we propose that the natural successor of the energy spectrum for an open-system LSM theorem is the entanglement spectrum.  

\textit{Statement and intuition of the open system LSM theorem.} 
Specifically, we propose an LSM-type theorem for open systems as follows: 
the low-lying spectrum of the entanglement Hamiltonian $\hat{K}$ 
either has a degenerate minimum or is gapless when the following two conditions are satisfied
\begin{enumerate}[leftmargin=*]
    \item SO$(3)$ symmetry and half-integer representation

    They are the key ingredients of the original LSM theorem and can be imposed on an open system. Namely, 
    we consider a half-integer spin chain coupled to a bath.  
    Both the translation and spin rotation symmetry in our setting is the so-called ``weak symmetry'' for the density matrix $\rho$ of the system~\cite{buca2012note}, i.e. 
    $
        \hat{U} \rho \hat{U}^\dag =\rho
    $
    or equivalently 
    \begin{equation}
    \label{eq:sym}
        \hat{U} \hat{K} \hat{U}^\dag =\hat{K},
    \end{equation}
    where $\hat{U}$ is either translation or spin rotation acting on the system. 
    When we have specific modeling of the bath, this weak symmetry condition is equivalent to requiring the total state of the full system, including the bath and system-bath coupling, to be invariant under the symmetry. 

    \item Short-range correlation
    
    In contrast to the long-range correlation implied by the closed system LSM theorem, we assume the spins in the open system are short-range correlated due to the coupling to the bath, i.e. $\langle\hat{O}_i\hat{O}_j\rangle-\langle\hat{O}_i\rangle\langle\hat{O}_j\rangle\sim e^{-|i-j|/\xi}$ for local operators $\hat{O}_i$ and $\hat{O}_j$ at site-$i$ and $j$ with a finite correlation length $\xi$. The expectation value $\langle \cdot \rangle$ is taken with respect to the system density matrix $\rho$. 
    We argue that this is the natural result when a system that satisfies the LSM theorem is coupled to a bath. 
    For instance, one scenario is that the system alone is enforced to be gapless by the LSM theorem, while the system--bath coupling is a relevant perturbation and opens up a finite gap $\Delta$. Hence, the system and bath together form a gapped ground state with $\xi\sim 1/\Delta$ roughly the inverse gap. Another scenario is that the system reaches thermal equilibrium with a thermal bath at inverse temperature $\beta=1/T$. In this case, $\xi\sim \beta$ is roughly the thermal length.
\end{enumerate}
 We compare this open system LSM theorem with the one for closed systems in Fig.~\ref{fig:ent_lsm}. 
 Given that ``entanglement'' serves a pivotal role in our proposal, we may alternatively call this open system LSM theorem as the {\it entanglement LSM theorem}.

Before discussing the details, we first present a special case to gain some physical intuitions on how the open system entanglement spectrum is related to the energy spectrum. 
Let us consider two spin-1/2 chains labeled by $a$ and $b$ with the same Hamiltonian $\hat{H}_{a,b}=\hat{H}$ that obeys the conditions for the original LSM theorem, e.g. we assume $\hat{H}=J\sum_i \hat{\bf S}_{i}\cdot \hat{\bf S}_{i+1}$ is the anti-ferromagnetic Heisenberg model. We then turn on a strong inter-chain anti-ferromagnetic coupling of strength $\Delta$ and treat $\hat{H}_{a,b}$ as perturbations. 
\begin{equation*}
\begin{tikzpicture}[scale=0.5,baseline={([yshift=-7pt]current bounding box.center)}]
\filldraw[fill=black] (0pt,0pt) circle (5pt);
\draw[->,>=stealth] (0pt,-10pt) -- (0pt,15pt);
\filldraw[fill=black] (0pt,-30pt) circle (5pt);
\draw[->,>=stealth] (0pt,-20pt) -- (0pt,-45pt);
\draw (0pt,-15pt) ellipse (8pt and 35pt);

\filldraw[fill=black] (40pt,0pt) circle (5pt);
\draw[->,>=stealth] (40pt,-10pt) -- (40pt,15pt);
\filldraw[fill=black] (40pt,-30pt) circle (5pt);
\draw[->,>=stealth] (40pt,-20pt) -- (40pt,-45pt);
\draw (40pt,-15pt) ellipse (8pt and 35pt);

\filldraw[fill=black] (80pt,0pt) circle (5pt);
\draw[->,>=stealth] (80pt,-10pt) -- (80pt,15pt);
\filldraw[fill=black] (80pt,-30pt) circle (5pt);
\draw[->,>=stealth] (80pt,-20pt) -- (80pt,-45pt);
\draw (80pt,-15pt) ellipse (8pt and 35pt);

\filldraw[fill=black] (120pt,0pt) circle (5pt);
\draw[->,>=stealth] (120pt,-10pt) -- (120pt,15pt);
\filldraw[fill=black] (120pt,-30pt) circle (5pt);
\draw[->,>=stealth] (120pt,-20pt) -- (120pt,-45pt);
\draw (120pt,-15pt) ellipse (8pt and 35pt);

\filldraw[fill=black] (160pt,0pt) circle (5pt);
\draw[->,>=stealth] (160pt,-10pt) -- (160pt,15pt);
\filldraw[fill=black] (160pt,-30pt) circle (5pt);
\draw[->,>=stealth] (160pt,-20pt) -- (160pt,-45pt);
\draw (160pt,-15pt) ellipse (8pt and 35pt);

\filldraw[fill=black] (200pt,0pt) circle (5pt);
\draw[->,>=stealth] (200pt,-10pt) -- (200pt,15pt);
\filldraw[fill=black] (200pt,-30pt) circle (5pt);
\draw[->,>=stealth] (200pt,-20pt) -- (200pt,-45pt);
\draw (200pt,-15pt) ellipse (8pt and 35pt);

\draw (-30pt,0pt) -- (230pt,0pt);
\draw (-30pt,-30pt) -- (230pt,-30pt);
\node at (-45pt,0pt) {$a$};
\node at (-45pt,-30pt) {$b$};
\end{tikzpicture}
\end{equation*}
To the zeroth order, each rung forms a singlet with the singlet--triplet energy gap $\Delta$ and $\Delta \gg J$ is much larger than the typical energy scale of $\hat{H}_{a,b}$. The wave function is a product of spin singlets
\begin{equation}
\ket{0}=\frac{1}{2^{L/2}}\bigotimes_{i=1}^L (\ket{\uparrow}_{i,a}\ket{\downarrow}_{i,b}-\ket{\downarrow}_{i,a}\ket{\uparrow}_{i,b}),
\label{eq:tfd}
\end{equation}
where $i$ labels sites. 
Now, including the $1$st order change of the wavefunction 
induced by the perturbation $\hat{H}_{a,b}$, we have 
\begin{equation}
\label{eq:pert}
\ket{\psi} \propto |0\rangle-\frac{1}{2\Delta}(\hat{H}_a+\hat{H}_b)|0\rangle\approx e^{-\beta (\hat{H}_a+\hat{H}_b)/2}|0\rangle,
\end{equation}
with $\beta=1/\Delta$. 
In the first step, we have simplified the $1$st order perturbation theory using the property that the Heisenberg Hamiltonian $\hat{H}_{a,b}$ flips exactly two singlets to triplets and therefore the energy difference is always $2\Delta$. Furthermore, one can show that the quantum state $\ket{0}$ can be re-written as
\begin{equation}
\ket{0}=\frac{1}{2^{L/2}}\sum_n\ket{E_n}_a\ket{\overline{E_n}}_b, 
\label{eq:tfd2}
\end{equation}
where $\{\ket{E_n}\}$ is the set of complete eigenstates of $\hat{H}$ and the overbar denotes time reversal.
With Eq.~\eqref{eq:tfd2}, it is straightforward to trace out the ``bath'' $b$ in state $|\psi \rangle$ and obtain the reduced density matrix for $a$, which is proportional to $e^{-\beta \hat{H}}$. Therefore, the entanglement spectrum satisfies the LSM theorem. 

\textit{Towards an information theoretical proof.} 
Now, we present a few steps that we believe are the keys towards a rigorous proof of the open system LSM theorem.

\begin{figure}
    \centering
    \includegraphics[width=0.48\textwidth]{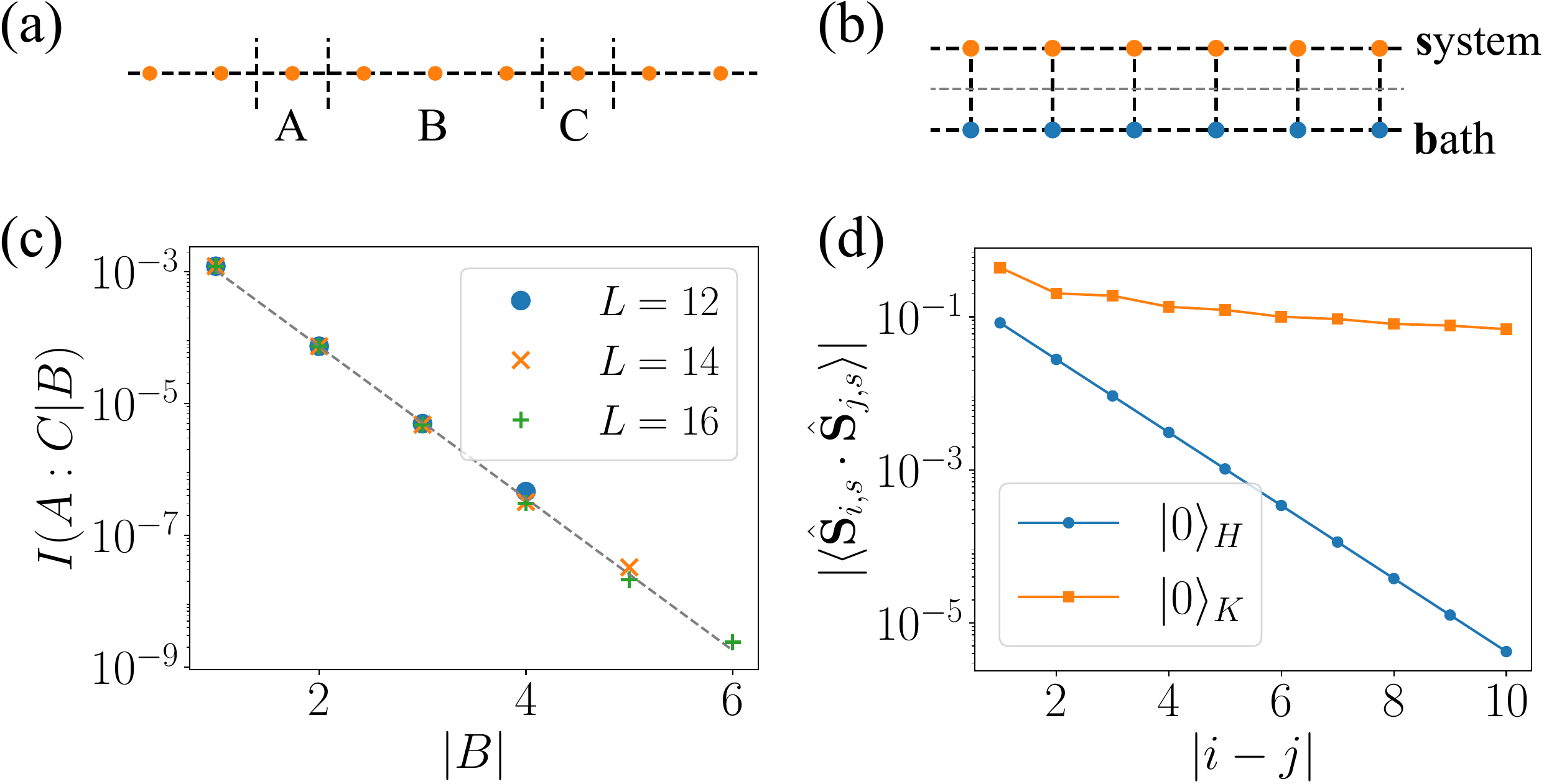}
    \caption{(a) The tripartition $A$, $B$ and $C$ of the spin chain as used in the proof. (b) Schematic of the model used in our numerical example. The upper chain with spin-$1/2$ denotes the system and the lower chain with spin-$3/2$ denotes the bath. They are coupled and the entire system respects translation and spin rotation symmetries. (c) The log--linear plot of the conditional mutual information $I(A:C|B)$ (see text for definition) as a function of the size of the subsystem $B$. (d) The log--linear plot of the spin--spin correlation function $|\langle{ \hat{\bf S}}_{i,s}\cdot{ \hat{\bf S}}_{j,s}\rangle|$ of system ``$s$'' as a function of the distance $|i-j|$ under the ground state $|0\rangle_H$ of the physical Hamiltonian $\hat{H}$ (blue dots) and the ground state $|0\rangle_K$ of the entanglement Hamiltonian $\hat{K}$ (orange squares). Note $|0\rangle_K$ lives in the Hilbert space of system $s$, which is a subspace of total Hilbert space where $|0\rangle_H$  lives. 
    \label{fig:numerics}}
\end{figure}

The first step is to argue that the state in question is a quantum approximate Markov state. This involves the notion of (conditional) mutual information. For two disjoint subsystems $A$ and $B$, the mutual information measuring the correlation between $A$ and $B$ is defined as $I(A:B)\equiv S_A+S_B-S_{AB}$, where $S$ is the von Neumann entropy of a subsystem labeled by its subscript. Then, the conditional mutual information between three disjoint subsystems $A$, $B$, and $C$ is defined as $I(A:C|B)\equiv I(A:BC)-I(A:B)=S_{AB}+S_{BC}-S_B-S_{ABC}$. The vanishing of the conditional mutual information (i.e. $I(A:C|B)=0$) roughly means that all correlations between $A$ and $C$ are bridged by $B$ and there is no direct correlation between $A$ and $C$. Consequently, a quantum Markov state is defined as the state satisfying $I(A:C|B)=0$. Now, consider the tripartition $A$, $B$, and $C$ shown in Fig.~\ref{fig:numerics}(a), where $B$ shields $A$ from $C$. When the total system is short-range correlated, we expect that $I(A:C|B)<\varepsilon$, where $\varepsilon$ is a small parameter controlled by $e^{-l/\xi}$ with $l$ the size of subsystem $B$ and $\xi$ a correlation length. Such a state is called a quantum $\varepsilon$-approximate Markov state \cite{QCMI,Kato2019,Svetlichnyy2022a,Svetlichnyy2022b}. 
While the exponential decay behavior is beyond rigorous proof so far, it is consistent with our numerics which will be discussed momentarily. 

The second step is to show that the entanglement Hamiltonian of a quantum approximate Markov state is local.
For a quantum Markov state with $I(A:C|B)=0$, it is known that the entanglement Hamiltonian is strictly local (finite range), more specifically, $\hat{K}_{ABC}=\hat{K}_{AB}+\hat{K}_{BC}-\hat{K}_B$ and therefore the entanglement Hamiltonian $\hat{K}$ for a large patch can be decomposed into a sum of the entanglement Hamiltonian for smaller patches~\cite{Petz:2002eql}. Now, for a quantum approximate Markov state with $\varepsilon \sim e^{-l/\xi}$, we conjecture that the corresponding entanglement Hamiltonian $\hat{K}$ is exponentially local, namely, the coefficients for non-local terms decay exponentially with the range of the term. 

The third step is to show that the spectral gap of $\hat{K}$ vanishes in the $L\rightarrow \infty$ limit. Following the idea of the original LSM theorem and admitting the exponential locality of $\hat{K}$, we can estimate the ``mismatch energy'' (the energy cost induced by the spin twist) as follows
\begin{equation} 
\begin{aligned}
    \langle \hat{K} \rangle_{\rm twist}- \langle \hat{K} \rangle_0 
   & \sim   L \int_0^\infty e^{-l/\xi} \left( 1- \cos \Big( \frac{2\pi l}{L}  \Big) \right) dl \\
     &\sim 
     \frac{4\pi^2 \xi^3}{L} - \frac{16 \pi^4 \xi^5}{L^3} + \ldots
\end{aligned}\label{eq:twist}
\end{equation}
Here the expectation value $\langle \cdot \rangle_0$ is taken on the ground state of $\hat{K}$ and $ \langle \hat{K} \rangle_{\rm twist} = \langle \hat{U}^{\rm twist \dagger} \hat{K} \hat{U}^{\rm twist} \rangle_0$ with the twist operator $\hat{U}^{\rm twist}$ defined in Eq.~\eqref{twist}. 
Since the  ground state is orthogonal to the twisted ground state that carries a different momentum as enforced by the non-trivial commutation relation between $\hat{U}^{\rm twist}$ and translation operator $\hat{T}$, 
the entanglement spectral gap is upper bounded by the above mismatch energy and therefore vanishes at $L\rightarrow \infty$, given $\xi$ a constant independent of $L$. 

\textit{Numerical Examples.} Now, we present numerical examples to illustrate the results more concretely~\footnote{The codes for our numerical calculations are available at https://github.com/chengshul/entLSM.}. As shown in Fig.~\ref{fig:numerics}(b), we let a spin-$1/2$ chain be the system in consideration and another spin-$3/2$ chain be the bath, which is reminiscent of a larger Hilbert space of the bath. The physical Hamiltonian of our first example is written as follows
\begin{equation}
\hat{H}=\sum_{i=1}^L J_1\big(\hat{\bf{S}}_{i}\cdot \hat{\bf{S}}_{i+1}+\frac{1}{3}(\hat{\bf{S}}_{i}\cdot \hat{\bf{S}}_{i+1})^2\big)+J_2 \hat{\bf{S}}_{i,s}\cdot \hat{\bf{S}}_{i,b},
\end{equation}
where $\hat{\bf{S}}_{i}=\hat{\bf{S}}_{i,s}+\hat{\bf{S}}_{i,b}$ is the sum of spin operators acting respectively on the system and bath, and $J_{1,2}>0$. $\hat{\bf{S}}_{i,s}$ and $\hat{\bf{S}}_{i,b}$ are the spin operators for the system and bath respectively. At each rung, the $J_2$-term is minimized by forming a spin triplet, rendering the $J_1$ term into the Affleck--Kennedy--Lieb--Tasaki (AKLT) Hamiltonian~\cite{Affleck1987}. 
This Hamiltonian has a gapped excitation spectrum and a unique ground state. 
We construct the model in this way such that its ground state wavefunction can be written down explicitly with the matrix product state representation. 
We let the total system stay in the ground state $|0\rangle_H$  of this Hamiltonian. 
Knowing the exact many-body wavefunction, we can obtain the reduced density matrix $\rho= \tr_{b} \left(|0\rangle \langle 0 |_H \right) $ of the system ``$s$'', and therefore, the entanglement Hamiltonian $\hat{K}=-\ln \rho$, to a fairly large system size.  

In Fig.~\ref{fig:numerics}(c), we first demonstrate the aforementioned statement on conditional mutual information, which is a crucial step in locality discussion. 
We plot $I(A:C|B)$ as a function of the system size of $B$ (the tripartition of $A$, $B$ and $C$ are shown in Fig.~\ref{fig:numerics}(a)), 
from which we see clearly an exponential decay as we discussed above. The decay length is found to be approximately $0.38$, which is of the same order as the correlation length of spin operators $\xi\approx 0.91$ on the ground state of the $\hat{H}$ as shown in Fig.~\ref{fig:numerics}(d) (blue dots). 
For comparison, we also compute the spin correlation function in the ground state $|0\rangle_K$ of the entanglement Hamiltonian $\hat{K}$ shown by the orange squares in Fig.~\ref{fig:numerics}(d), which confirms the implications of our entanglement LSM theorem that the state $|0\rangle_K$ cannot be short range correlated. 

\begin{figure}
    \centering
    \includegraphics[width=0.4\textwidth]{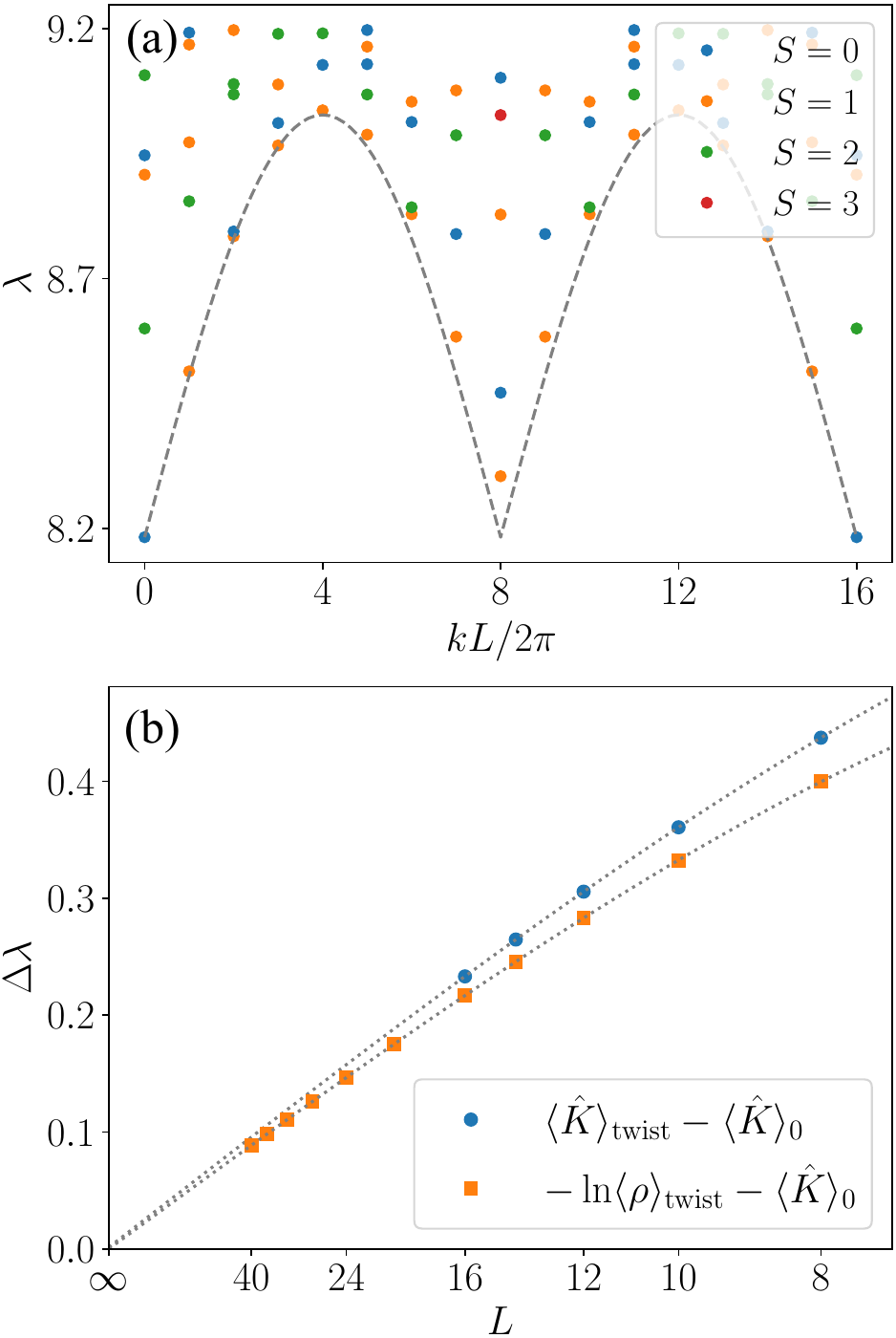}
    \caption{(a) The entanglement spectrum with momentum and total spin resolved. The dashed line is given by $\lambda= v|\sin{k}|+\lambda_0$, with $v$ and $\lambda_0$ obtained from the two lowest points at $k=0,2\pi/L$. The calculation is done for $L=16$. (b) $\langle \hat{K}\rangle_\text{twist}-\langle \hat{K}\rangle_0$ (blue points) and $-\ln \langle \rho\rangle_\text{twist}-\langle \hat{K}\rangle_0$ (orange squares) are plotted as a function of the total system length $L$. Both quantities are upper bounds of the entanglement spectral gap $\Delta \lambda$ of the entanglement Hamiltonian $\hat{K}$. The dotted lines are fitting curves with $c_0+c_1/L+c_3/L^3$. 
    The $1/L^2$ term and all higher even order terms vanish due to the inversion symmetry generated by $\exp (i\pi\sum_jS_j^x)$. 
    The fitting results are 
    $(c_0,c_1,c_3)=(0.0014,3.8,-19)$ and $=(0.00013,3.6,-23)$ for $\langle \hat{K}\rangle_\text{twist}-\langle \hat{K}\rangle_0$ and $-\ln\langle \rho \rangle_{\rm twist}-\langle \hat{K}\rangle_0$ respectively.  
    Note the ratio of coefficients $-c_1/c_3<3/\pi^2$ is consistent with the expectation from expanding a ``mismatch'' energy
    $\propto 1-\cos \left( 2\pi r/L  \right)$, where $r\in \mathbb{Z} $ is the range of the coupling. The ratio reaches the maximum $3/\pi^2$ when only nearest neighbor terms are involved. 
    \label{fig:numerics2}
    }
\end{figure}

Next, we calculate the spectral properties of $\hat{K}$ 
to show the vanishing of the spectral gap in the large system size limit. 
First, for a total system size $L=16$, we compute the entire entanglement spectrum, as shown in Fig.~\ref{fig:numerics2}(a). This spectrum is consistent with a linear gapless spectrum \cite{newpaper}. 
Then, we focus on the entanglement spectral gap (denoted as $\Delta \lambda$) estimation as discussed above in Eq.~\eqref{eq:twist}. We numerically calculate both $\langle\hat{K}\rangle_\mathrm{twist}$ and $\langle\hat{K}\rangle_0$ and fit the difference against $c_0+c_1/L+c_3/L^3$. 
A less intuitive but numerically 
more feasible quantity is  $-\ln\langle \rho \rangle_{\rm twist}-\langle \hat{K}\rangle_0$ where 
$\langle \rho \rangle_{\rm twist}=\langle\hat{U}^{\mathrm{twist}\dagger} \rho\hat{U}^\mathrm{twist}\rangle_0$. 
Using the convexity of the function $-\ln(x)$, it can be shown that 
this quantity provides a tighter bound as follows
\begin{equation}
    \Delta \lambda \leq -\ln\langle \rho \rangle_{\rm twist}-\langle \hat{K}\rangle_0
    \leq \langle \hat{K}\rangle_\text{twist}-\langle \hat{K}\rangle_0 .
\end{equation}
We demonstrate the numerical results of both quantities in Fig.~\ref{fig:numerics2}(b). 
As shown in the data, the finite size scaling supports the statement that the entanglement spectral gap $ \Delta \lambda$ vanishes at the thermodynamic limit $L\rightarrow \infty$. 
Similar numerical results have also been reported previously for two coupled spin-$1/2$ chains~\cite{Poilblanc2010}, which can be regarded as another example supporting our result.

\begin{figure}
    \centering
    \includegraphics[width=0.48\textwidth]{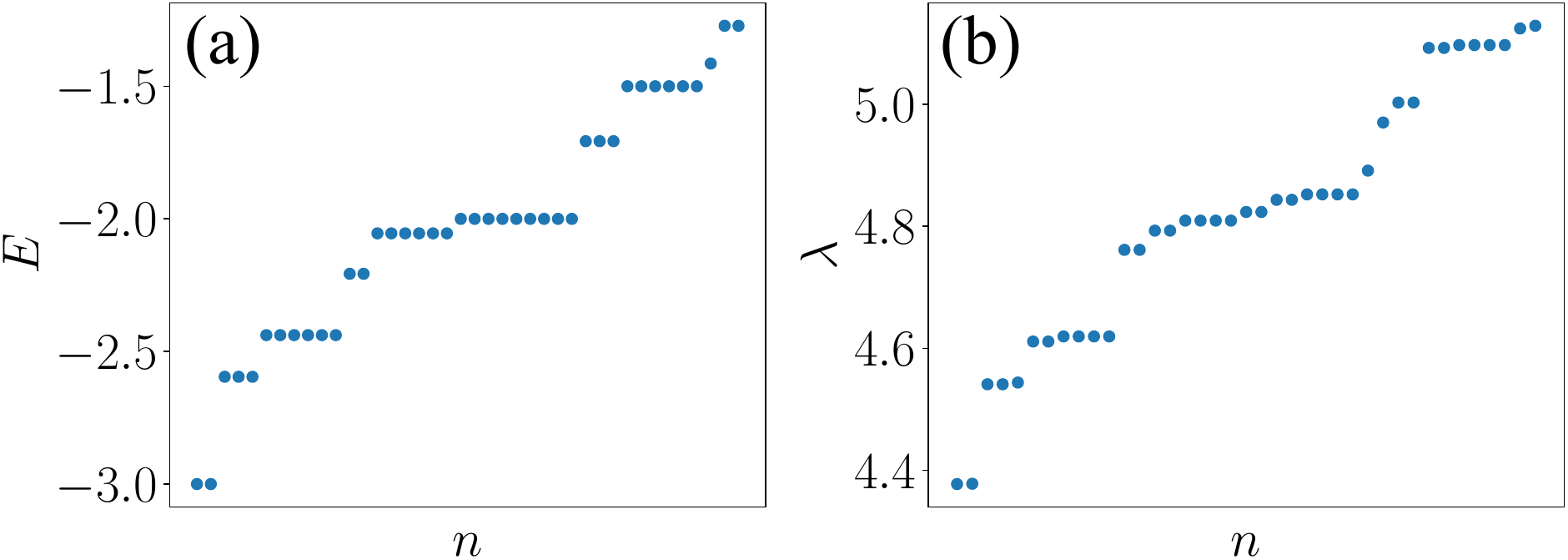}
    \caption{(a) The energy spectrum of the original Majumdar--Ghosh model. (b) The entanglement spectrum of the  Majumdar--Ghosh model coupled to a spin-3/2 bath, Eq.~\eqref{eq:mg}. In both cases, we have the system length $L=8$.
    \label{fig:ssb}}
\end{figure}

The second model we consider is a spin-1/2 Majumdar--Ghosh (MG) chain~\cite{Majumdar1969} coupled to a spin-3/2 bath, where we expect the entanglement Hamiltonian to exhibit spontaneous symmetry breaking. The model reads
\begin{equation}\label{eq:mg}
    \begin{aligned}
\hat{H}=\sum_{i=1}^L ~~&J_1(\hat{\bf{S}}_{i,s}\cdot \hat{\bf S}_{i+1,s}+\frac{1}{2}\hat{\bf{S}}_{i,s}\cdot \hat{\bf {S}}_{i+2,s})\\
+ ~&J_2 \hat{\bf{S}}_{i,s}\cdot \hat{\bf{S}}_{i,b}+D(\hat{S}_{i,s}^z+\hat{S}_{i,b}^z)^2.
    \end{aligned}
\end{equation}
Here we have introduced an anisotropic coupling $D$ and set $J_1=J_2=D=1$ in the numerics. Note that we now have O(2) instead of SO(3) global symmetry, which too suffices for LSM theorem \cite{Affleck1986}. The ground state reduces to a product state of $|S=1, S_z=0\rangle$ in the $J_1\rightarrow0$ limit, and remains trivially gapped for $J_1=1$ which we have checked numerically. In Fig.~\ref{fig:ssb} we plot the energy spectrum of the original MG model and the entanglement spectrum. In both cases, a doubly degenerate ground state energy/entanglement eigenvalue characteristic of SSB is observed.

\textit{Summary and Outlook.} For a spin chain where the original LSM theorem was valid, coupling to the bath overturns the prediction of the LSM theorem. 
Such coupling can turn the local operators short-range correlated. However, precisely because of the short-ranged correlation, the entanglement Hamiltonian acquires locality, a key ingredient for the revival of the LSM theorem in the entanglement spectrum of an open quantum system.

In this letter, we show that familiar topological constraints on the energy spectrum can be extended to the entanglement spectrum of an open system that is entangled with the environment. While compared with existing literature relating the non-trivial topology and the entanglement spectrum such as the Li--Haldane conjecture \cite{Li2008} where a one-dimensional entanglement cut is performed on a two-dimensional topological state, the entanglement cut discussed in our work is in the same dimension as the system. It will also be interesting to systematically explore the connections between these two scenarios \cite{qiGeneralRelationshipEntanglement2012,hsieh2014bulk,gu2016holographic}.

In closed systems, the developments of the LSM-type theorem have revealed its rich content,
including the generalization to more sophisticated on-site and space groups \cite{chen2011classification,Parameswaran13, Watanabe_2015,Po_2017}, symmetry-protected and enriched topological systems \cite{Zaletel_2015,Cheng2016,Cho2017,huang2017buildinga,lu2017liebschultzmattis,Yang_2018,jian2018liebschultzmattis,Metlitski2018, Ogata2019, Else2020,Lu_2020,jiang2021generalizeda,ye2022topological}, 
and fermionic systems \cite{fidkowski2018surface,Cheng_2019,zheng2022unconventional}. These studies have shown an enormous impact on topological and correlated quantum phases. 
Following the progress made in this work, 
all these generalizations could also be discussed in the content of entanglement Hamiltonian for open systems. 
In parallel to the closed systems, these studies may also strongly impact the understanding of topology and correlation in open quantum many-body systems, an area on the frontier of the current theoretical research \cite{bardynTopologyDissipation2013,dasbiswas2018topological,wanjura2020topological,gneiting2022unraveling,leefmans2022topological,bao2023mixedstate,fan2023diagnosticsa,mao2023dissipation,ma2023topological,su2023conformal} and practically relevant to the quantum simulators in the NISQ era \cite{preskill2018quantum,altman2021quantum}.

\textit{Acknowledgement}. 
We thank Zhen Bi, 
Meng Cheng, 
Chao-Ming Jian, 
Shenghan Jiang,  
Chong Wang and 
Chao Yin 
for discussions. 
This work is supported by 
the Innovation Program for Quantum Science and Technology 2021ZD0302005, 
the Beijing Outstanding Young Scholar Program, 
the XPLORER Prize, NSFC Grant No.~12042505, 
China Postdoctoral Science Foundation Grant No.~2022M711868
and Tsinghua University Dushi Program. 
C.L. is also supported by the Chinese International Postdoctoral Exchange Fellowship Program and the Shuimu Tsinghua Scholar Program at Tsinghua University. 
The DMRG calculations are performed using the ITensor library (v0.3) \cite{ITensor}.

\bibliography{biblio}

\end{document}